# Nanoscale control of the metal-insulator transition at LAO/KTO (110) and LAO/KTO (111) interfaces

Muqing Yu[1,2], Changjiang Liu[3], Dengyu Yang[1,2], Xi Yan[3], Qianheng Du[3], Dillon D. Fong[3], Anand Bhattacharya[3], Patrick Irvin[1,2], Jeremy Levy[1,2]

[1]*Department of Physics and Astronomy, University of Pittsburgh, Pittsburgh, PA 15260, USA*
[2]*Pittsburgh Quantum Institute, Pittsburgh, PA, 15260, USA*
[3]*Materials Science Division, Argonne National Laboratory, Lemont, IL 60439, USA*

**Abstract**

Recent reports of superconductivity at $KTaO_3$ (KTO) (110) and (111) interfaces have sparked intense interest due to the relatively high critical temperature as well as other properties that distinguish this system from the more extensively studied $SrTiO_3$ (STO)-based heterostructures. Here we report nanoscale control of the metal-to-insulator transition at the $LaAlO_3$/KTO (110) and (111) interfaces. Devices are created using two distinct methods previously developed for STO-based heterostructures: (1) conductive atomic-force microscopy lithography and (2) ultra-low-voltage electron-beam lithography. At low temperatures, KTO-based devices show superconductivity that is tunable by an applied back gate. A nanowire device shows single-electron-transistor (SET) behavior. These reconfigurable methods of creating nanoscale devices in KTO-based heterostructures offer new avenues for investigating mechanisms of superconductivity as well as development of quantum devices that incorporate strong spin-orbit interactions, superconducting behavior, and nanoscale dimensions.

The rich physical properties of complex-oxide heterostructures and the paradigm of semiconductor nanoelectronics have begun to overlap, resulting in the new field of correlated nanoelectronics[1]. Significant advances have been made with SrTiO$_3$ (STO)-based heterostructures[2], which host an exceptionally wide range of gate-tunable physical properties including superconductivity[3–5], magnetism[6–8], ferroelectricity[9–12], and ferroelasticity[13–16]. Superconductivity in STO was discovered more than half a century ago[17], and there is still no consensus about the underlying mechanism[18].

The discovery of superconductivity in heterostructures based on KTaO$_3$ (KTO) represents a breakthrough as a "sibling" system that share many similarities with STO (Ref. [19–21]). Liu et al. reported superconductivity at EuO/KTO(111) and LaAlO$_3$ (LAO)/KTO(111) interface with a superconducting transition temperature $T_c \approx 2$ K (Ref. [19]). Interfacial superconductivity was also observed with LAO/KTO (110) heterostructures with $T_c \approx 0.9$ K (Ref. [21]). The band structure of KTO is similar to STO except that the conduction band of KTO comes from the $5d$ orbital of Ta rather than $3d$ orbital of Ti[22,23]. Like STO, KTO is a "quantum paraelectric" with a nearly diverging permittivity that saturates at low temperatures [24], presumably due to quantum fluctuations of the ferroelectric soft mode. It was known previously that superconductivity at the KTO (001) surface, with carriers induced by electrolyte gating, is exceptionally weak, with $T_c \approx$ 50 mK and with very low critical current densities[25]. LAO/KTO (001) heterostructures host 2D electron gases (2DEGs), but superconductivity was not reported[19].

Insight into correlated electron phenomena, such as superconductivity at oxide interfaces, can be obtained by the creation of nanoscale devices. Among the various methods, conductive atomic-force microscopy (c-AFM) lithography enables sub-10-nm devices to be "sketched" and "erased"[26], resulting in a wide range of mesoscopic devices whose properties can help constrain physical mechanisms and potentially find use in future quantum device applications. Experiments with LAO/STO devices have created quasi-2D, 1D, and 0D structures that provide insight into the nature of the superconducting state[27–29]. Recently, Yang et al. reported an alternate way to create electronic devices at the LAO/STO interface using ultra-low-voltage electron-beam lithography (ULV-EBL)[30]. This technique offers significantly faster writing speed compared with c-AFM lithography, while maintaining comparable resolution and device quality. This technique provides a new avenue for producing significantly more complex devices and for integrating complex-oxide heterostructures with other quantum materials.

Here we adapt the c-AFM lithography and ULV-EBL techniques to patterning the LAO/KTO (110) and LAO/KTO (111) interfaces. The conductivity of the KTO interface depends on the thickness of the LAO layer, with an observed critical thickness $t_c \approx 3$ nm (Ref. [31]). A LAO/KTO heterostructure is prepared by depositing a 3-nm-thick amorphous LAO film on KTO (110) single crystal by pulsed laser deposition (see Methods). Electrical contact to LAO/KTO interface is fabricated by $Ar^+$ milling, followed by sputter deposition of Ti/Au; a second layer of Ti/Au on top is added to create pads for wire bonding (see Methods). The c-AFM lithography setup is illustrated in Figure 1A. The LAO/KTO (110) interface is initially insulating with $G <1$ nS conductance measured between any pair of electrodes. Figure 1B shows an AFM image of one of the LAO/KTO (110) "canvases." Green areas indicate regions that are to be scanned with a positive bias and red areas indicates regions to be scanned with negative bias. Once the main channel is completed, a sharp four-terminal conductance jump $\Delta G_{4T} \approx 300$ nS is observed (Figure 1C), indicating the creation of conducting path between the two pairs of leads 1,2 and 7,8 (see Methods). The c-AFM writing with LAO/KTO (110) is reversible: When the tip is negatively biased ($V_{tip} = -8$ V) and moved across the existing wire along the red path in Figure1B inset (with tip velocity $v_{tip} = 50$ nm/s), a steep drop in the conductance is observed (Figure 1D), which can be used to estimate the width of the nanowire (see Methods section and Ref. [26]). The conductance and the width of the nanowire (~20 nm) is comparable to wires created at the LAO/STO interface[26], indicating a similar mechanism for creating and erasing conducting channels. Most likely, the surface charging of the LAO surface[32,33], created by the voltage-biased c-AFM tip, is responsible for nanoscale control of the insulator-to-metal transition at LAO/KTO (110) interface. The nature of the surface charges has not been determined, but the most likely sources are ionized oxygen vacancies and water-derived protons[34,35].

We also demonstrate creating conducting channels at LAO/KTO (110) interface with ULV-EBL. The experimental setup of ULV-EBL is shown in Figure 1E. Six aluminum wire bonds are affixed onto the LAO surface in a circular arrangement to form a ULV-EBL canvas (Figure 1F) that is initially insulating with conductance between any two pairs of wire bonds $G < 1$ nS). During ULV-EBL (see Methods), a voltage $V_1$ is sourced from electrode 1 and current $I_{15}$ is measured through electrode 5 along with voltage $V_{87}$ across electrode pair (8,7) (Figure 1F). Thus, two-terminal conductance is given by $G_{2T} = I_{15}/V_1$, while the four-terminal conductance is given by $G_{4T} = I_{15}/V_{87}$. At first the blue regions are exposed with a total areal dose of 30 $\mu C/cm^2$ (Figure 1F). Then the main channel (Figure 1F, green, length = 60 μm, width = 15 μm) is thrice exposed with an aerial dose of 10 μC/cm², causing three conductance jumps to be observed (Figure 1G). The four-terminal conductance reaches $G_{4T} = 6$ μS, indicating the creation of conducting channels at the LAO/KTO interface ("Device A").

Device A is transferred to a dilution refrigerator immediately after exposure by ULV-EBL. The decay of KTO-based devices is slowed significantly when the device is stored in vacuum, similar to LAO/STO devices[32], and conductance decay is completely arrested when the device is cooled to low temperatures. The time when the sample is exposed to air is less than 5 minutes (Figure S3). Device A remains metallic down to the lowest temperatures measured (~50 mK) (Figure 2A). At $T = 50$ mK and $B = 0$ T, a flat region in four-terminal current-voltage measurement ($I - V$ measurement) is observed for small current values (Figure 2B). Based on the $I - V$ curve, the four-terminal differential resistance $dV/dI$ drops sharply from ~7 kΩ to ~120 Ω when the bias current falls below a critical current $I_c$, defined as the averaged values where the differential resistance reaches a sharp maximum (Figure 2B). For Device A, $I_c = 18$ nA, corresponding to a critical current density $j_c = 12$ uA/cm.

Figure 2C shows $dV/dI$ versus $I$ curve acquired at $B = 0$ T and temperatures in the range $T = 50$ to 330 mK. Figure 2D shows $dV/dI$ versus $I$ curves acquired at $T = 60$ mK with different out-of-plane magnetic fields. Figure 2E is the plot of differential resistance at zero bias ($dV/dI$ at $I = 0$ nA) versus temperature, which yields $T_c = 156$ mK, defined to be where the resistance reaches half of the normal state resistance $R_N$. The upper critical field $B_{c2}$ can be similarly defined to be the field at which the zero-bias differential resistance reaches half of its normal-state value. Here we find (Figure 2F) $B_{c2} = 140$ Oe.

Since the dielectric constant of KTO grows to $\epsilon \approx 4{,}500$ at low temperatures[24], applied voltages to the back of the KTO substrate ($V_{bg}$) are able to modulate the carrier density of interface devices. Hall measurements are performed on Device A at temperatures above the superconducting phase (at $T = 0.7$ K) for various back gate voltages, while sourcing from electrode 1 to electrode 5 and measuring voltage $V_{82}$ across electrode pair 8 and 2 (Figure 3A). The measured sheet carrier density $n_s$ and electron mobility $\mu$ are summarized in Figure 3B. Through back gating, the carrier density can be tuned from $5.6 \times 10^{12}$/cm^2 to $8.1 \times 10^{12}$/cm^2. The average carrier density is approximately one order of magnitude lower than what has been previously reported for KTO (110) or KTO (111)[19,21]. The low carrier density may attribute to the low $T_c$ and $B_{c2}$ of this device. The mobility $\mu$ increases significantly when the gate voltage is increased from $-20$ V to $+5$ V, and saturates when $V_{bg} > 5$ V. Figure 3D shows an intensity plot of $dV/dI$ as a function of $V_{bg}$ and $I$ at $T = 50$ mK and $B = 0$ T. The critical current increases monotonically with increasing $V_{bg}$, and saturates at $V_{bg} \approx 5$ V. As shown by the normalized resistance versus temperature (Figure 3D), $T_c$ also increases monotonically with increasing $n_{2D}$, providing evidence that the device is in the underdoped regime. Regions that have not been exposed by ULV-EBL remain insulating up to the largest back gate voltages applied, as verified by two-terminal $I - V$ measurements between electrode 5 and 4 at $V_{bg} = 50$ V (Figure S4).

Here we also demonstrate a quasi-one-dimensional conducting channel at the LAO/KTO (110) interface created by ULV-EBL. Prior to ULV-EBL writing, the sample is left in ambient conditions for one week to allow previously-written conducting patterns to decay, thus restoring the insulating interface (< 1 nS between electrodes). The layout of the device ("Device B") is shown in Figure 4A. The blue-shaded region is exposed three times with aerial dose of 10 $\mu C/cm^2$, while the yellow region is exposed six times with 10 $\mu C/cm^2$ dose. The main channel (green, Figure 4A inset) consists of a single line that is 16 $\mu$m long, exposed 40 times with linear dose of 50 pC/cm. The resolution of ULV-EBL writing is reported to be comparable to c-AFM lithography[30], so we estimate that this nanowire device has a width comparable to what is measured by c-AFM lithography ~20 nm.

Differential conductance $dI/dV$ versus $V_{bg}$, measured at $T = 50$ mK, show a sequence of diamond-shaped low-conductance regions (Figure 4B). The zero-bias conductance (Figure 4C) shows a series of conductance peaks as a function of $V_{bg}$, with a factor of 2.5 difference between adjacent peaks and bottoms. The overall characteristic strongly resembles the behavior of a single-electron transistor, except that there is no artificially defined island. Rather, it is more likely that the disorder potential has produced a landscape where resonant tunneling through a quasi-zero-dimensional island dominates the overall transport. Further experiments are needed in order to effectively create, manipulate, and utilize quantum dots at the LAO/KTO interface and to deliberately create single-electron transistors and ballistic quantum channels, similar to what has been demonstrated in LAO/STO[29,36]. Furthermore, charging of these islands may indicate a state of electron pairing without superconductivity[28]. Here, no evidence of such a pairing transition has been established.

C-AFM lithography and ULV-EBL have also been tested with the LAO/KTO (111) heterostructures. An intrinsically insulating 6 nm LAO/KTO (111) interface was created by PLD (see Methods). Using both c-AFM lithography (Figure S5A, B, C) and ULV-EBL (Figure S5D, E), conducting structures can be created. However, this device exhibited insulating behavior below $T \approx 85$ K (Figure S5F), which precluded further investigation of low-temperature properties such as superconductivity.

In summary, we have demonstrated nanoscale control of the metal-insulator transition in LAO/KTO (110) and LAO/KTO (111) heterostructures, using two different techniques: c-AFM lithography and ULV-EBL. The devices which survive to milli-Kelvin temperatures exhibit key signatures of superconductivity that are tunable with a back-gate voltage. A quasi-one-dimensional conducting channel is conductive to milli-Kelvin temperatures and exhibits signatures of single-electron-transistor (SET) behavior.

In terms of device patterning, each of these two techniques has its own strengths. ULV-EBL is able to create large 2D patterns quickly, while c-AFM lithography has the ability to erase locally, which is suitable for making nanoscale barriers. Both techniques offer a wealth of insights into the nature of the superconducting state, based on prior experiments performed with LAO/STO. In addition, the unique differences—the lack of a bulk structural phase transition, and the significantly larger spin-orbit interaction strength—will help improve our understanding of how superconductivity arises with such dilute carrier concentrations and potentially create new families of quantum devices that can exploit these properties.

**Methods**

*LAO film growth*
Amorphous LAO thin films were prepared on the as received 5 × 5 mm KTO substrates using pulsed laser deposition (PLD) method. The substrates were pre-annealed in situ at 500 °C at the base pressure of our PLD system, in the low $10^{-6}$ Torr range, for 30 mins before growth. During the deposition, temperature and pressure were kept the same, and a 248-nm KrF excimer laser was used with 2 Hz repetition rate and 1.5 J/cm² laser fluence. A commercially available 10 × 10 mm single crystal LAO (001) wafer was used as the target. After growth, the samples were cooled down to room temperature at a rate of 5 °C/min in the same pressure as during growth. The thickness of the amorphous LAO film was calibrated using x-ray reflectivity and the surface morphology was measured using atomic force microscopy.

*Patterning of metal electrodes for c-AFM lithography*
LAO/KTO samples were patterned with standard photolithography using AZ4110 photoresist. The exposed regions after developing first underwent Argon ion milling (100 V acceleration, 10 mA emission) for 15 minutes. The etched regions were filled by 5.5 nm titanium and 36.5 nm gold to make electrical contact to the heterostructure interface. Both ion milling and metal deposition were done in a Plassys 8-pocket e-gun evaporator with built-in ion gun. Excessive metal was lifted off in acetone. Finally, bonding pads were created by depositing 4 nm titanium followed by 50 nm gold. Figure S1 shows a typical c-AFM lithography canvas.

*Conductive-AFM lithography*
The conductive atomic force microscope lithography was performed using an Asylum Research MFP-3D AFM. The DC writing voltage was applied to the c-AFM tip (highly doped silicon) through a 10 MΩ tip resistor. As shown in Figure 1B, the AFM tip first sketches the four solid "virtual electrodes" from electrodes 1, 2, 7, 8 toward the device region, using a bias voltage $V_{tip} = 20$ V. After sketching these four solid conducting shapes, the main channel (green path) is sketched with $V_{tip} = 16$ V. During the c-AFM lithography process, an ac voltage is sourced at Lead 2 ($V_2 = V_0 \cos(2\pi f t)$, where $t$ is time, $V_0 = 100$ mV, $f = 13$ Hz). Current $I_{27}$ through Lead 7 and voltage $V_{18}$ across Leads 1 and 8 (Figure 1B) are measured using a lock-in amplifier, and the four-terminal conductance is given by $G_{4T} = I_{27}/V_{18}$.

*Estimation of the width of the nanowire*
The width of a conductive wire is extracted by fitting the conductance drop (Figure 1D) with to a parameterized function [26]:

$$G(x) = \frac{a_1}{2} * \left(1 - a_2(v_{tip}t - x_0)\right)\left(1 - \tanh\left(\frac{v_{tip}t - x_0}{W}\right)\right) + a_3$$

where $a_{1-3}$ are fitting parameters, $x_0$ represents the center position of the wire, and $W$ is the width of the wire, extracted to be $W = 19.9$ nm.

*Ultra low voltage electron beam lithography (ULV-EBL)*
The ULV-EBL was performed in a Zeiss Gemini 450 scanning electron microscope (with Zeiss Gemini 2 column). Lithography step was controlled by Raith Elphy Plus system. Metal markers were deposited near four corners of LAO/KTO samples for focusing and alignment. Unwanted exposures within ULV-EBL canvases are avoided. The electron beam exposes a certain area with an acceleration voltage of 500 V, which is enough for the electrons to reach the LAO/KTO interface (Figure S2). The electron beam current is measured to be 18 pA, and the write field is set to be 650 μm × 650 μm.

*Low-temperature transport measurement*
Low-temperature transport measurements were performed in a Quantum Design PPMS dilution refrigerator. Source voltages were applied by a 24-bit digital/analog converter National Instruments PXI-4461, which can also simultaneously perform 24-bit analog/digital conversion. The drain current and the voltages were measured after amplification by a Krohn-Hite Model 7008 Multi-channel Pre-amplifier.

**Acknowledgements**

**Funding:** J.L. acknowledges support from the DOE-QIS program (DE-SC0022277). J.L. and P.I. acknowledge support from NSF (PHY-1913034). All work at Argonne was supported by the US Department of Energy, Office of Science, Basic Energy Sciences, Materials Sciences and Engineering Division. The use of facilities at the Center for Nanoscale Materials and the Advanced Photon Source, both Office of Science user facilities, was supported by the US Department of Energy, Basic Energy Sciences under Contract No. DE-AC02-06CH11357.

**Author contributions:** J. L. and A. B. supervised the experiments. C. L., X. Y., Q. D. and D. F. performed the sample growth. M. Y., D. Y. and P. I. performed c-AFM lithography, ULV-EBL and transport measurements at low temperature. M. Y. analyzed the data. All authors discussed the results and commented on the manuscript.

**Competing interests:** The authors declare that they have no competing interests.

**Data Availability:** Data for all graphs presented in this paper are available from the corresponding author upon reasonable request.

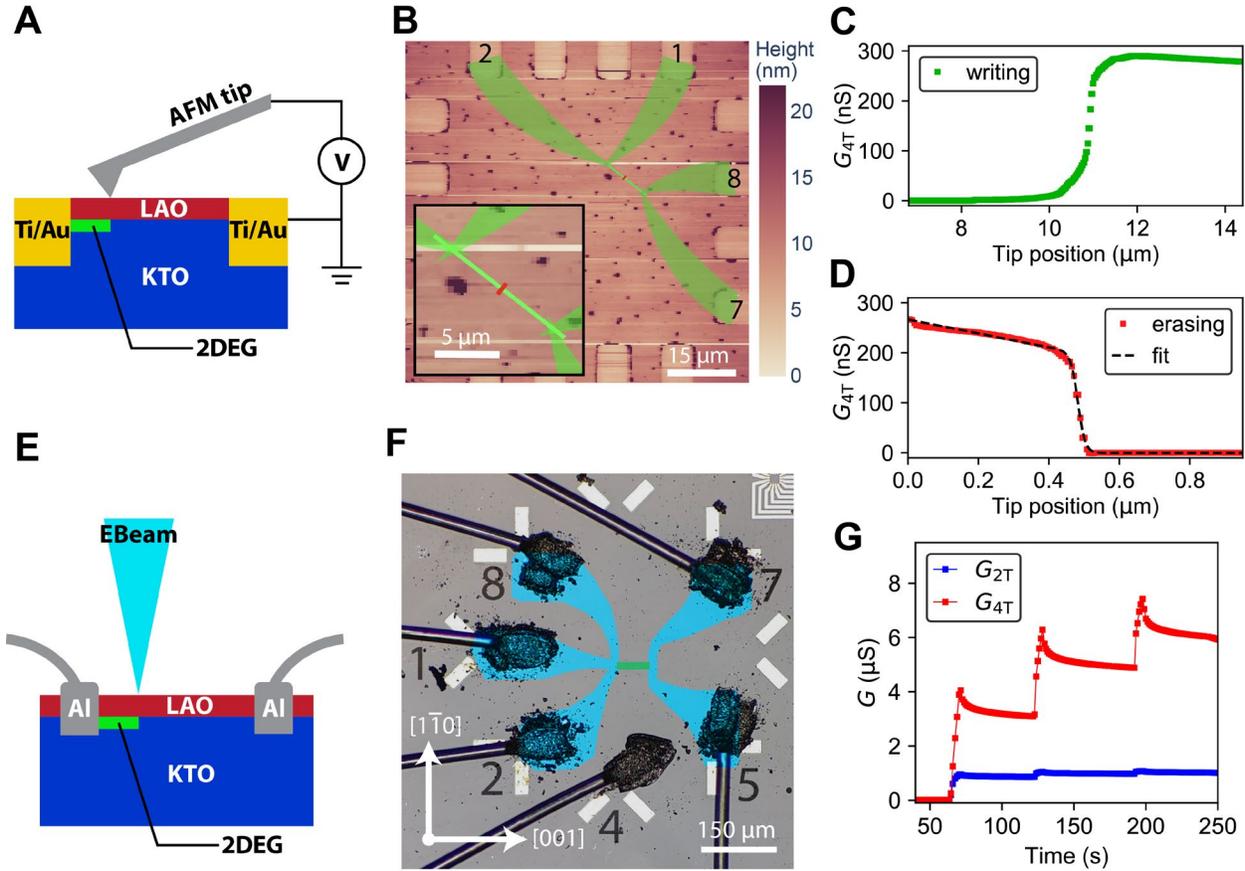

**Figure 1. Reconfigurable patterning of LAO/KTO heterostructures.** (A) C-AFM method in which a biased conductive AFM tip is scanned over the LAO surface. (B) Top view of c-AFM pattern, where green areas depict regions where $V_{\text{tip}} > 0$, and red regions depict areas where $V_{\text{tip}} < 0$. (C,D) Combined experiment where a single nanowire is sketched between leads 1,2 and 7,8. Subsequent erasure yields an estimate of the nanowire width $w \approx 20$ nm. (E) ULV-EBL writing method in which ~500 V electrons are focused on a pattern, also resulting in local changes to the conductance at the LAO/KTO interface. (F) Top view of a patterned area. Interface contacts are obtained by Al wire bonds, and blue regions represent "virtual electrodes". Electrical conductance between leads 8,1,2 and 5,7, $G_{4T} = dI_{15}/dV_{87}$ is achieved by a final rectangular pattern (width $w = 15\ \mu m$, length $l = 60\ \mu m$), indicated in green. (G) Measured four-terminal and two-terminal conductance $G$ as the last rectangular region is patterned. Each conductance jump corresponds with one exposure with 10 $\mu C/cm^2$.

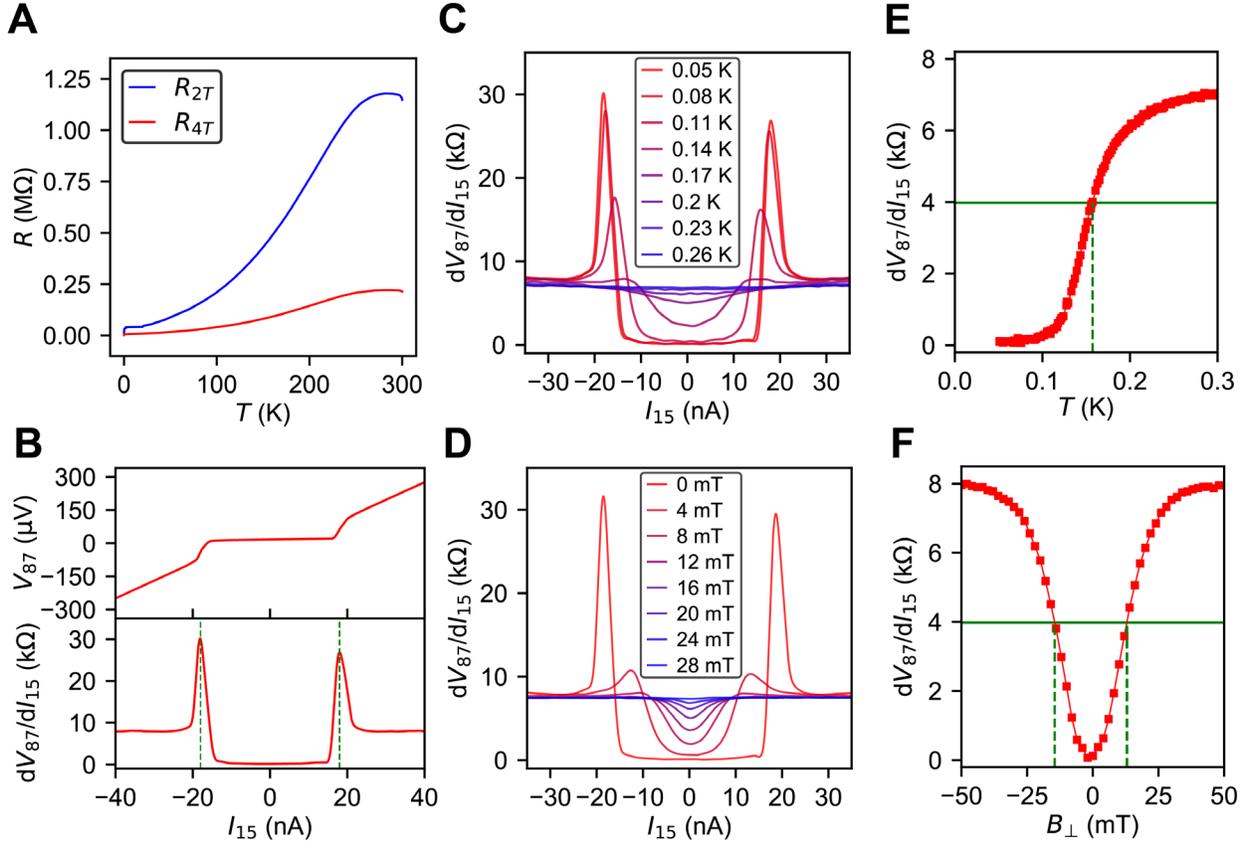

Figure 2. Temperature-dependent magnetotransport of LAO/KTO (110) Device A patterned with ULV-EBL. (A) Two-terminal and four-terminal resistance versus temperature. (B) $I-V$ and $dV/dI$ versus $I$ at $T = 50$ mK and $B = 0$ T. Green dashed lines indicate the critical current. (C) $dV/dI$ vs $I$ over a temperature range 50 mK $\leq T \leq$ 260 mK. (D) $dV/dI$ versus $I$ for 0 T $\leq B \leq$ 0.028 T acquired at $T = 60$ mK. (E) Zero-bias ($I = 0$) differential resistance vs temperature. Green dashed line indicates temperature at which resistance drops by 50%. (F) Zero-bias ($I = 0$) differential resistance vs magnetic field at $T = 60$ mK. Green dashed line indicates magnetic field at which resistance drops by 50%.

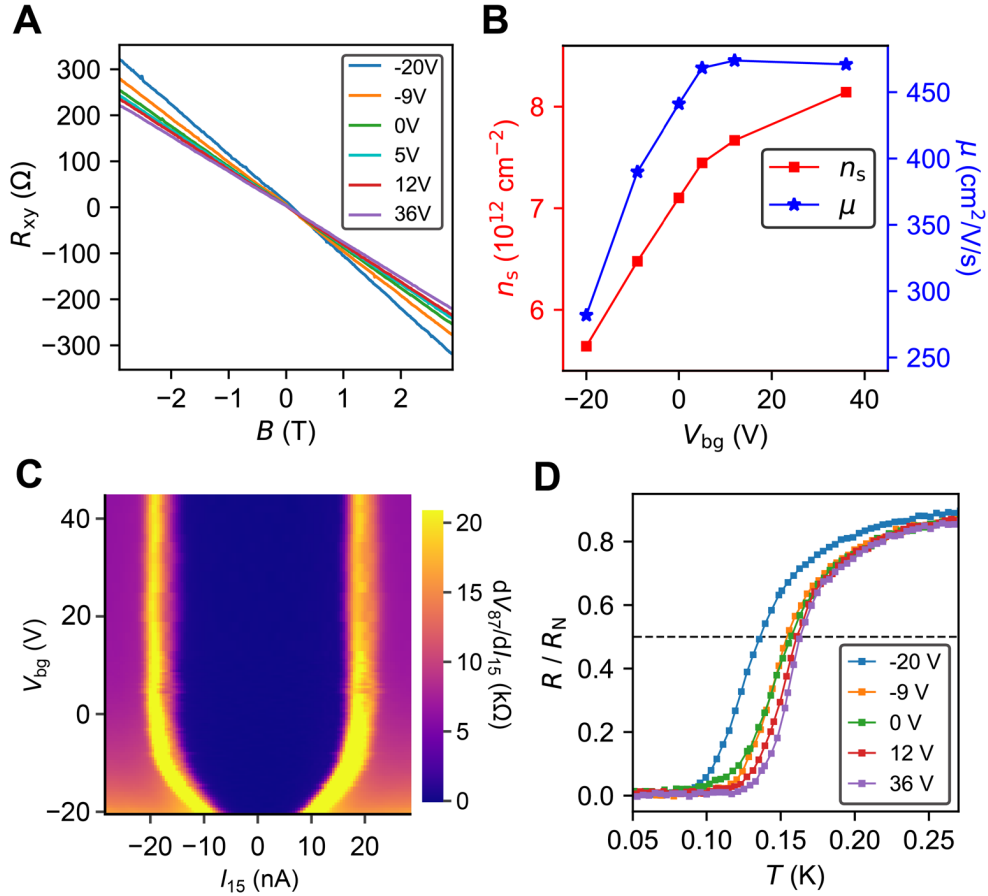

Figure 3. Back gate ($V_{bg}$)-dependent transport for Device A. (A) Hall resistance versus magnetic field. (B) Hall carrier density $n$ and mobility $\mu$. (C) Differential resistance versus $I$ and $V_{bg}$, at $T = 50$ mK and $B = 0$ T. (D) Normalized resistance versus temperature.

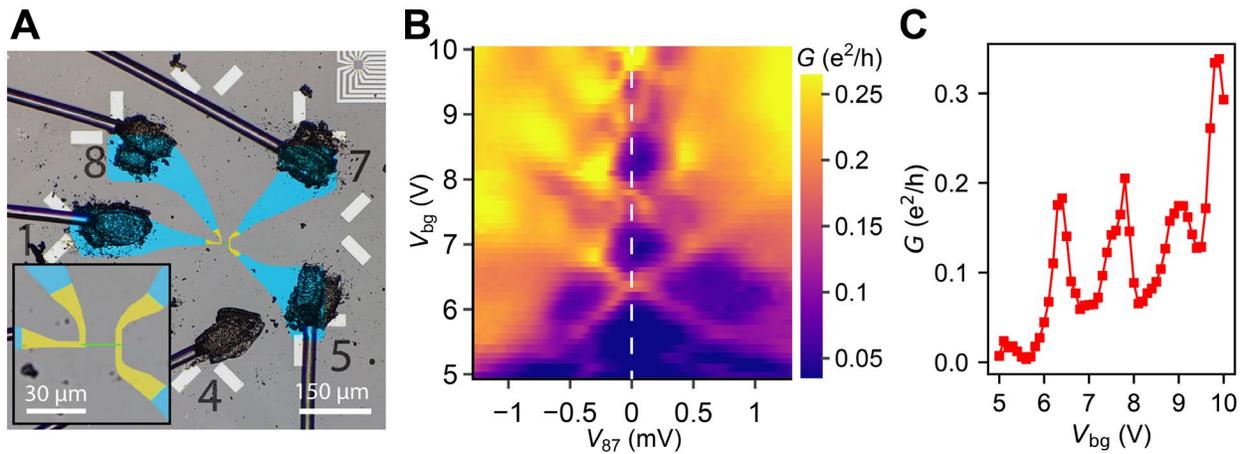

Figure 4. Transport measurements on a quasi-1D nanowire Device B. (A) Nanowire device design. (B) Differential conductance $dI/dV$ versus $V$ and $V_{bg}$, showing evidence for discrete charging of islands. (C) Linecut showing resonant tunneling peaks.

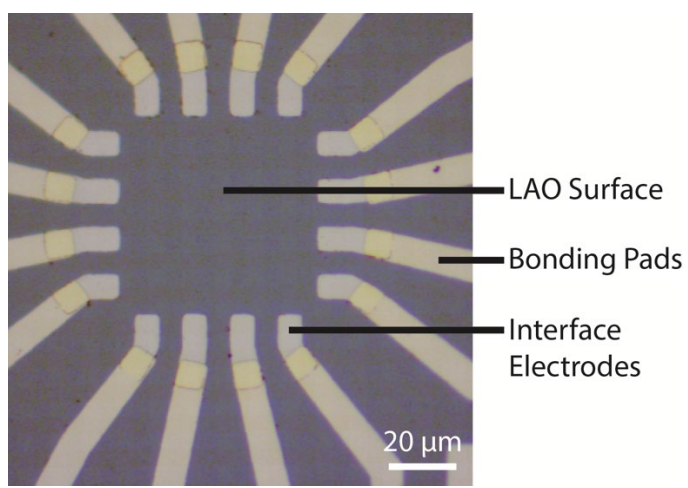

Figure S1. Optical image of a c-AFM lithography canvas.

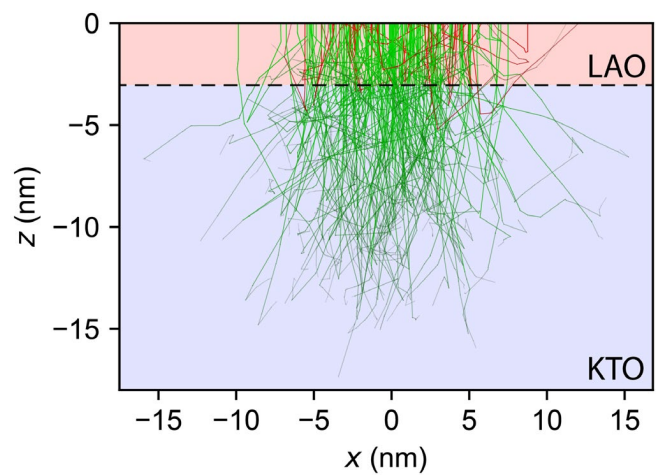

Figure S2. Monte Carlo simulation of electron trajectories in 3 nm LAO/KTO done by CASINO. Acceleration voltage is set to be 500 V. Red lines represent the paths of backscattered electrons. Green lines represent the paths of injected electrons.

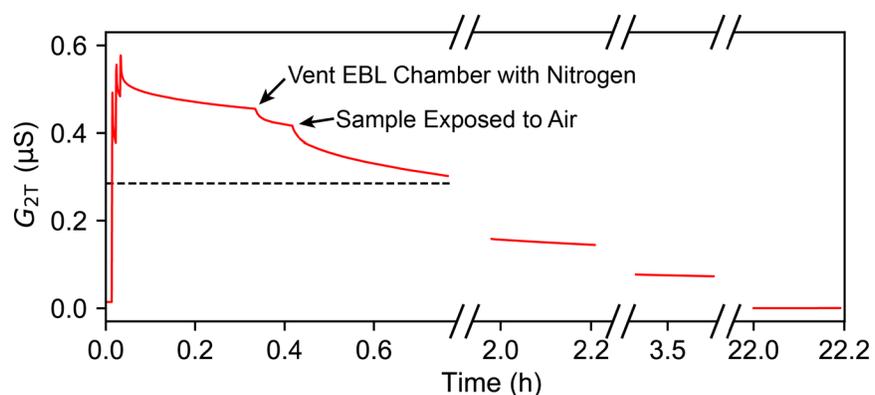

Figure S3. Decay of two-terminal conductance versus time after ULV-EBL writing. The device measured here has the same shape and is exposed with the same dose as Device A depicted in Figure 1F. Instead of being transferred into a Dilution Refrigerator, this device is left in the EBL chamber after the ULB-EBL exposure, with its conductance continuously monitored. Two-terminal conductance is determined by $G_{2T} = I_{15}/V_1$. Black dashed line indicates 50% value of the initial conductance jump. If the device is left to decay in ambient environment, its conductance will reach zero after 22 hours. Device A was transferred into dilution refrigerator within 0.1 hours of exposure to air, which did not substantially affect its conductance.

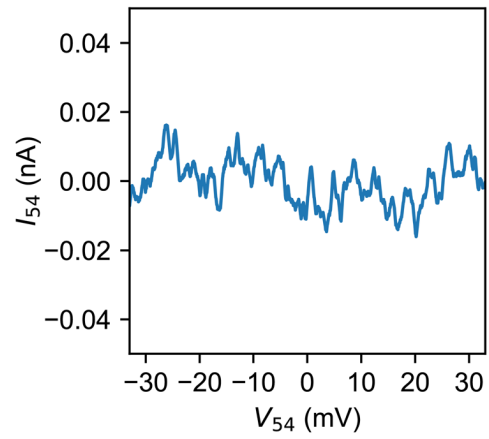

Figure S4. *I-V* curve of Device A measured between electrodes 4 and 5 ($T = 50$ mK and $V_{\text{bg}} = +50$ V).

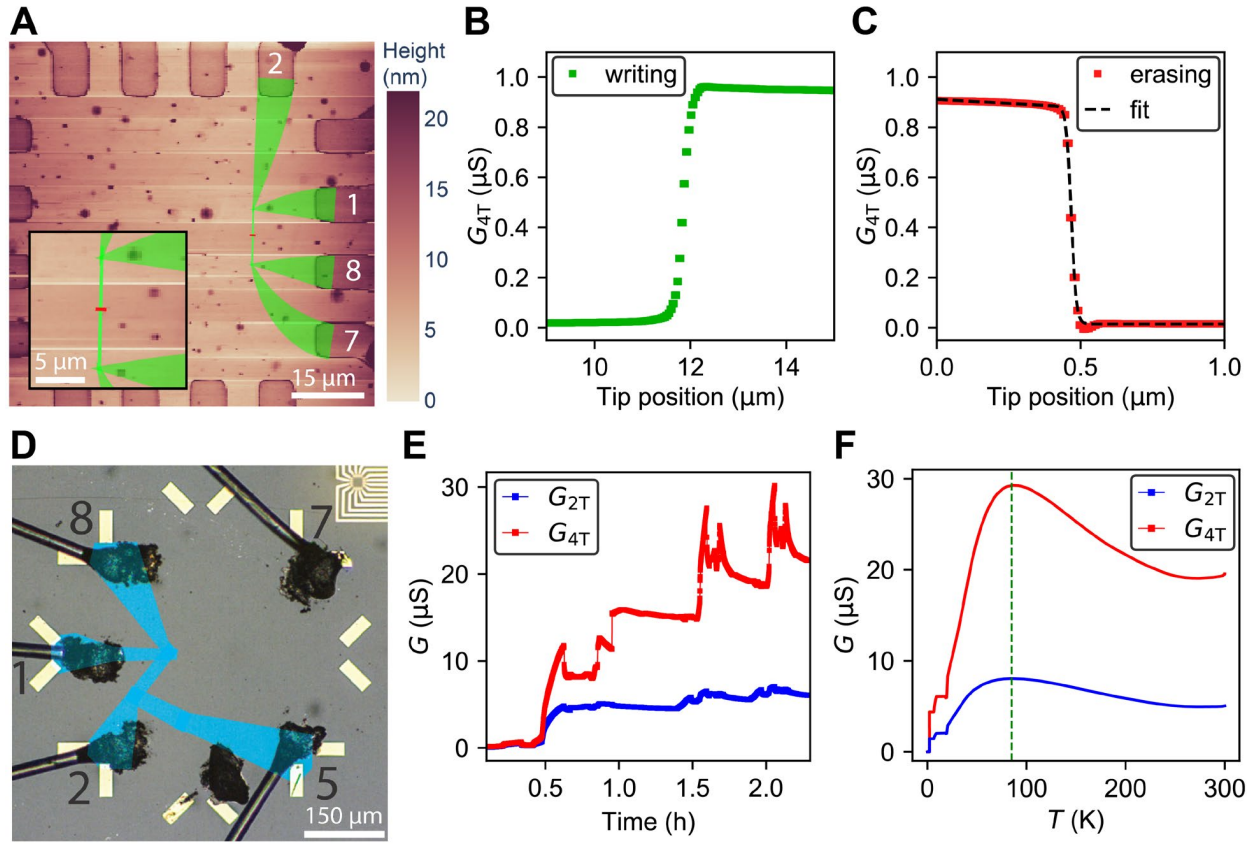

Figure S5. C-AFM lithography and ULV-EBL patterning on 6 nm LAO / KTO (111) sample. (A) Top view of c-AFM pattern, where green areas depict regions where $V_{\text{tip}} > 0$, and red regions depict areas where $V_{\text{tip}} < 0$. (B) Four-terminal conductance jump during the writing of main channel (green path) in (A). (C) Four-terminal conductance decreases back to 0 during the cut of main channel along the red path in (A). Curve fit (dashed curve) gives the width of main channel ~17 nm. (D) Device patterned by ULV-EBL. The blue region is exposed by a total areal dose of 230 µC/cm². During ULV-EBL, a voltage $V_8$ is sourced from electrode 8 and current $I_{82}$ is measured through electrode 2 along with voltage $V_{15}$ across electrode pair (1,5) (E) Measured two-terminal and four-terminal conductance: $G_{2T} = I_{82}/V_8$ and $G_{4T} = I_{82}/V_{15}$ during exposure. (F) Conductance versus temperature of the device depicted in (D). The device starts to become insulating at ~ 85 K (green dashed line).